# Personalizing Smart Home Privacy Protection With Individuals' Regulatory Focus: Would You Preserve or Enhance Your Information Privacy?

Reza Ghaiumy Anaraky
rg4598@nyu.edu
New York University
USA

Yao Li
yao.li@ucf.edu
University of Central Florida
USA

Hichang Cho
hichang_cho@nus.edu.sg
National University of Singapore
Singapore

Danny Yuxing Huang
dhuang@nyu.edu
New York University
USA

Kaileigh A. Byrne
kaileib@clemson.edu
Clemson University
USA

Bart Knijnenburg
bartk@clemson.edu
Clemson University
USA

Oded Nov
onov@nyu.edu
New York University
USA

## ABSTRACT

In this study, we explore the effectiveness of persuasive messages endorsing the adoption of a privacy protection technology (IoT Inspector) tailored to individuals' regulatory focus (promotion or prevention). We explore if and how regulatory fit (i.e., tuning the goal-pursuit mechanism to individuals' internal regulatory focus) can increase persuasion and adoption. We conducted a between-subject experiment (N = 236) presenting participants with the IoT Inspector in gain ("Privacy Enhancing Technology"—PET) or loss ("Privacy Preserving Technology"—PPT) framing. Results show that the effect of regulatory fit on adoption is mediated by trust and privacy calculus processes: prevention-focused users who read the PPT message trust the tool more. Furthermore, privacy calculus favors using the tool when promotion-focused individuals read the PET message. We discuss the contribution of understanding the cognitive mechanisms behind regulatory fit in privacy decision-making to support privacy protection.

## 1 INTRODUCTION

Studies show that people can be reluctant to manage their online privacy and security; for example, they may be unwilling to explore the privacy settings on a social media account to prevent an unfavorable disclosure [73], or be uninterested in adopting protective technologies, such as virtual private networks that can address security vulnerabilities [81]. Therefore, finding ways to motivate users to adopt protective technologies is an important challenge in information security and privacy management [71, 75, 79, 81, 96, 99, 100].

One way to increase individuals' motivation to take action is to provide them with personalized messages that specifically appeal to them [26, 51, 83, 106]. For example, a message highlighting excitement and social rewards is found to be more persuasive for extroverts considering the purchase of a new cell phone [51]. In this work, we study the efficacy of using personalized messages to persuade individuals to adopt a privacy-protection technology. While previous privacy literature has shown that using different message framing may lead to various privacy-protective behaviors [4, 7, 90], no work has examined whether such framing should be tailored to different users, i.e., users with different regulatory foci. To that end, we draw on regulatory focus theory to motivate our research questions and design [44, 49]. Regulatory focus theory suggests that people's goal orientation is a trait variable (i.e., similar to personality traits) and can be promotion-focused or prevention-focused [44, 49]. Individuals with a high promotion focus have a high desire for growth [49]. Those with a high prevention focus desire safety and direct their efforts to prevent unfavorable outcomes [49]. This makes regulatory focus especially relevant to privacy management [20, 63], as privacy management can be equally viewed as either a function of effective privacy risk *prevention* or as the *promotion* of privacy protection.

When one's internal regulatory focus matches one's goal-pursuit strategy, there is a *regulatory fit* [44]. Research shows that with regulatory fit, individuals show more positive attitudes towards the task [33] and are more likely to fulfill it [64]. In this paper, we explore whether regulatory fit can enhance the efficacy of personalized persuasive messages on individuals' adoption of a privacy-protection technology, and seek to explore the underlying mechanisms through which regulatory fit may increase individuals' pursuit of a goal:

- **RQ**: Does tailoring a persuasive message encouraging individuals to adopt a privacy-protection technology to their internal regulatory focus (i.e., regulatory fit) increase adoption behavior? What are the mechanisms through which regulatory fit increases this adoption behavior?

Privacy literature depicts privacy calculus and trust as prominent mediators of privacy decisions [16, 27, 28, 36]. To answer our research question and understand the mechanisms through which regulatory fit may increase adoption behavior, we study both privacy calculus and trust as potential mediators of the effects of regulatory fit on privacy behavior. Consequently, we designed a between-subject experiment and studied the adoption of "IoT Inspector" as a privacy-protection technology. IoT Inspector is a tool that can help smart-device users monitor the network communications of their smart devices [53]. Using this tool, users can view the domains with which their devices communicate and the size of these communications in bits. We used IoT Inspector as it is a widely adopted open-source tool that helps non-expert smart-device users explore their privacy [53]. We introduced IoT Inspector either with a gain framing of *Privacy **Enhancing** Technology (PET)* or a loss framing of *Privacy **Preserving** Technology (PPT)* to 236 participants recruited through Prolific, a crowd-sourcing platform. After reading the introductory message (either PET or PPT), participants answered survey questions regarding their initial impressions of the tool and were given a chance to actually download and use the IoT Inspector.

Our findings show how *privacy calculus* that users make for using vs. not using the IoT Inspector and their *trust* in the tool can change based on the message framing (PET vs. PPT) and the user's regulatory focus. Specifically, we found that regulatory fit for promotion-focused individuals can induce a more positive privacy calculus towards the product: when we present individuals who have a high promotion focus with the "privacy-enhancing technology" message framing, they have more thoughts in favor of using the tool. On the other hand, regulatory fit for prevention-focused users can induce trust: when we present individuals who have a high prevention focus with the "privacy-preserving technology" message framing, they trust the technology more and, in turn, are more likely to install the tool.

To the best of our knowledge, this is the first study to examine the persuasive effect of regulatory fit in the privacy decision-making domain. Our work contributes to the theoretical understanding of regulatory fit as we explore its underlying mechanisms by studying potential mediations by privacy calculus and trust. Furthermore, our findings have practical implications for marketing a product to potential customers. Finally, this work can inform policy design and help encourage individuals to adopt privacy-protection technologies.

## 2 LITERATURE REVIEW AND HYPOTHESIS DEVELOPMENT

### 2.1 Persuading to Adopt Privacy Measures

Scholars have explored ways to persuade individuals to protect their privacy. For example, they have discovered that communicating the purpose of data collection can persuade individuals to disclose their data [62], and using password meters can increase the likelihood of them choosing stronger passwords [19]. However, these persuading mechanisms do not consistently lead to the desired outcome. Huh et al.[54] surveyed LinkedIn users who received password reset appeals from LinkedIn. They found that users are reluctant to reset their passwords, since after several weeks, only around 46% of email recipients changed their passwords. They highlight the ineffectiveness of current persuasive mechanisms that are used by companies. In another study, Egelman et al. [30] studied the efficacy of messages encouraging users to choose stronger passwords. They found that while the messages led to choosing stronger passwords in a hypothetical scenario, such interventions did not consistently lead to stronger passwords in real scenarios. In another field study, Ghaiumy Anaraky et al. [8] showed that persuasive messages that encourage people to automatically tag themselves in Facebook photos would result in lower persuasion and tagging behaviors if the default setting is opt-in. These mixed results show that persuading individuals to take privacy protection measures is a complex topic.

In order to uncover the efficacy of persuasive messages, it is crucial to consider two points: first, it is important to study if and how the effect of persuasive messages on behavior is mediated by the key relevant variables (e.g., privacy calculus and trust in the privacy domain). Second, personal characteristics such as regulatory focus are central to persuasion [20, 69, 115]. It is likely that users with different regulatory foci react differently to persuasive messages. Currently, no work has considered designing privacy persuasive messages with respect to individuals' regulatory focus and testing the mediating role of privacy calculus and trust.

### 2.2 The Role of Privacy Calculus and Trust in Privacy Decision Making

In this section, we briefly explain how privacy calculus and trust can influence privacy decisions. Then, we discuss how loss- and gain-framed messages can influence privacy calculus, trust, and behaviors.

#### 2.2.1 *Privacy Calculus.*

Privacy calculus theory suggests that privacy decisions involve making a trade-off between the risks and benefits associated with a decision [24]. Many studies have adopted the privacy-calculus model to explore the dynamics behind privacy decisions [17, 43, 55, 94, 108]. For example, Li et al. [68] studied the adoption of wearable healthcare devices based on the privacy-calculus model. They showed that when users decide to adopt wearables, they consider the benefits they may gain from using the device as well as the risks involved with using the device (e.g., having to disclose personal information). Consequently, their findings suggest that if the benefits of using the wearable outweigh the privacy risks, people are more likely to opt for using the wearable. On the other hand, if the risks outweigh the benefits, they are less likely to use the wearable. In another study, Jalali et al. [57] explored the adoption of Internet of Things devices as a function of their risks and rewards. Their work suggests that while low levels of risk and high reward would result in the highest adoption rates, high levels of risk and low reward would result in no adoption. Based on the privacy-calculus literature, we hypothesize the following:

- **H1**: Individuals whose privacy calculus is more strongly in favor of adopting the privacy-protection technology are more likely to download it.

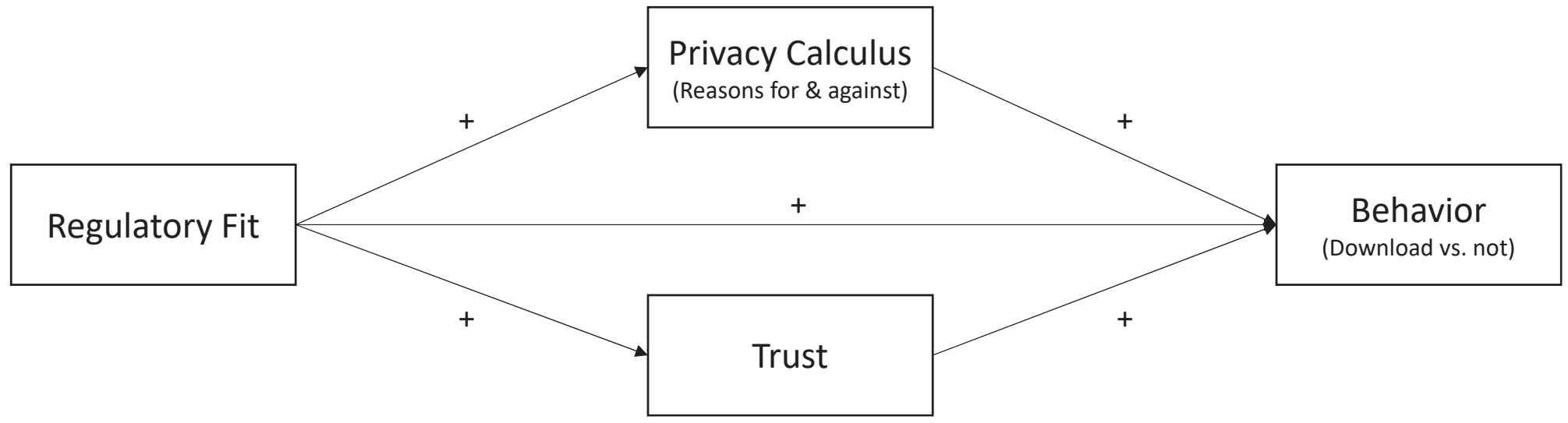

**Figure 1: Our Conceptual model through which we explore the effects of regulatory fit on the evaluations of the product (privacy calculus), trust, and privacy behaviors (download behavior).**

While privacy calculus made significant contributions to the privacy literature, we know that it has shortcomings. The underlying assumption in the privacy-calculus model is that humans make deliberate trade-offs between the risks and benefits of decisions. However, research shows that this assumption does not always hold as individuals have limited cognitive resources when making privacy decisions [16, 28]. Consequently, many scholars provide evidence suggesting that privacy calculus is not the sole mechanism involved in privacy decision-making and that individuals use mental shortcuts to fast-track their privacy decisions [1, 2, 8, 36, 101]. In the following, we explain how trust acts as a mental shortcut for privacy decisions.

*2.2.2 Trust.* Several studies highlight the important role of trust in privacy decisions [27, 28, 66]. Lewicki [66] presents trust as one of the decision shortcuts since trusting an entity can reduce individuals' sensitivity to information. This, in turn, reduces the complexity of the decision-making [66, 67]. Consequently, a trusted brand name or a trusted authority plays a significant role in users' decisions about revealing personal information online [103]. In another study, Dinev and Hart [27] showed that a high trust can overwrite the perceived risks associated with information disclosure to an e-commerce website. Furthermore, trust is found to be a predictor in adopting virtual private networks [81]. In line with this literature, we pose the following hypothesis:

- **H2**: Individuals who trust the privacy-protection technology more, are more likely to adopt it.

## 2.3 Tailoring Loss and Gain—Framed Messages to Regulatory Focus

Presenting the same information with different wordings can result in various behavioral outcomes [105]. This is called the framing effect and is studied in different disciplines such as Psychology [13] and Economics [21]. Several studies found the framing effect to be an effective means of persuading individuals to take an action [40, 41, 89]. These studies often present the scenario in terms of gains—presenting favorable consequences of taking an action, or losses—presenting unfavorable consequences of not taking an action. For example, Peng et al. [89] studied the efficacy of gain and loss-framed messages in persuading individuals to take COVID-19 vaccines. They found a loss-framed message explaining the negative consequences of not being vaccinated more persuasive than a gain-framed message explaining the benefits of getting vaccinated. Tversky and Kahneman [105] suggest loss aversion as the reason behind the framing effect. This means that losses loom greater than gains, and when individuals see the loss-framed message, they would be more inclined to take action than when they see a gain-framed message [58].

Additionally, research highlights the effects of message framing on various attitudes [12, 35, 77, 112]. This effect is broadly explored in health communication literature. For example, when people read a loss-framed message—cautioning people about the unfavorable consequences of not taking a cancer preventative measure (vs. a gain-framed message—explaining the benefits of taking a cancer preventative measure), they show more positive attitudes about the preventative behavior [77]. Similarly, a loss-framed message about a cancer screening test can increase a reader's perceived cancer susceptibility more than a gain-framed message [35].

The effect of message framing has also been studied in security and privacy research. Most studies have found more persuasive effects from loss-framed messages compared to gain-framed messages. For instance, Ma and Birrell found that users who saw the cookie banner with negative framing (i.e., "degrade your experience") were more likely to accept cookies, compared to positive framing (i.e., "improve your experience") [72]. Similarly, Qu et al. found that showing disadvantages in the message framing (i.e., "your account is at risk") can be useful to nudge participants [90]. Acquisti et al. found that users tend to sacrifice privacy when asked to "pay $2 to protect their privacy" but tend to protect privacy when

asked to "give away privacy for $2" [3, 39]. Adjerid et al. found that a privacy notice suggesting an "increase" in privacy protection also results in increased disclosures, while a privacy notice suggesting a "decrease" in privacy protection elicits decreased disclosures [4]. However, some studies did not find a significant effect of framed messages. For example, DeGiulio et al. did not find a difference between messages that emphasize the benefits of allowing tracking vs. messages that emphasize the potential negative consequences of opting out of tracking [25].

With one notable exception [12], previous studies have not examined the underlying mechanisms that cause the effects of gain and loss framing. The change of framing is a heuristic manipulation [28] that can manifest differently in terms of cognitive and emotional appraisals. For example, the default effect—another prominent heuristic effect—suggests that people are more likely to proceed with the option that is pre-selected by default. Some have argued that this is a cognitively mediated effect (i.e., users may perceive a pre-selected option as an implicit endorsement and decide to select it accordingly [29, 56]), while others have argued that the default effect is an affect-based manifestation stimuli (i.e., influencing users' emotional appraisal of the proposed options [34]). Similarly, we explore the effects of loss and gain framing heuristic on privacy calculus, which constitutes a cognitive evaluation of the input signal, and trust, which encompasses an emotional appraisal. The next hypothesis explores the effect of framing on our dependent variables:

- **H3**: A loss-framed (vs. gain-framed) message leads to having a) the privacy calculus in favor of using the privacy-protection technology, b) a higher trust for the privacy-protection technology, and c) higher adoption of the privacy-protection technology.

Additionally, studies have shown that the effectiveness of a message framing depends on the characteristics of the audience [91, 109]. Privacy management involves both the effective prevention of privacy risks and the promotion of privacy enhancement. As such, regulatory focus is a relevant trait in examining both approach- and avoidance-related goal orientations pertaining to privacy.

*2.3.1 Regulatory Focus.* The regulatory focus theory [44, 49] suggests that individuals have two distinct motivational systems for pursuing goals: promotion and prevention systems. These systems originate from different fundamental human needs and seek different outcomes (goals). The promotion system is derived from the desire for growth and nurturance. Individuals with a high promotion focus are concerned with having positive outcomes over the absence of positive outcomes (having gains over non-gains). The prevention system is derived from the need for safety and security. Individuals with a high prevention focus emphasize avoiding negative outcomes over the presence of such outcomes (having non-losses over losses) [49]. In addition, promotion and prevention systems are independent, such that an individual can have high promotion and high prevention or low promotion and low prevention foci [49].

To explain promotion and prevention foci further, Higgins [46, 47] discusses the distinct ways these systems construe their end-goal state. Higgins considers "0" as the status quo state. People with a strong promotion focus consider the state of "+1" as the gain or success status. Therefore, not achieving this gain (i.e., maintaining the "0" status quo) looms as a loss for these individuals. On the contrary, individuals with a strong prevention focus consider the maintenance of the "0"—the status quo— and not going below it a success, and the "-1" state a failure.

Regulatory focus theory has been used in the Human-Computer Interaction (HCI) literature to promote user experience in human-robot interactions [5, 23, 31], privacy decision-making [20, 63], and human interactions with virtual agents [32]. Le et al. [63] used regulatory focus theory to study individuals' privacy decision-making in a mobile payment application. They found that users with a high prevention focus are more cautious and have a lower intention to disclose personal information. However, promotion-focused users are more likely to disclose information if the disclosure scenario serves their goals. Cho et al. [20] showed that people with a high promotion focus have a more positive attitude about managing their online privacy preference on a social media platform and perceive this task as less effortful. In the following, we discuss our approach to leveraging regulatory focus in personalizing privacy interventions.

*2.3.2 Regulatory Fit.* Regulatory fit happens when an individual's motivational orientation (i.e., promotion or prevention foci) matches their goal-pursuit strategy [44]. When individuals experience regulatory fit, they are more likely to be persuaded to take an action [64] such as purchasing a product [11] or getting tested to see if they have a disease and need treatment [86]. For example, Werth and Foerster studied how regulatory fit affects consumers' purchasing behavior [109]. They created two versions of a car advertisement. In one of the advertisements, they focused on safety aspects (aspects important for individuals with a high prevention focus), while in the other advertisement, they emphasized comfort (aspects important to individuals with a high promotion focus). They found that when the advertisement aligns with the consumer's regulatory focus (i.e., when there is regulatory fit), the consumer expresses more positive opinions about the product than when the advertisement is incompatible with the consumer's regulatory focus [109].

These behavioral effects of regulatory fit may stem from its capacity to promote positive attitudes about the task. For example, regulatory fit can promote perceived enjoyment [33] and performance [45, 65]. Overall, with a regulatory fit, individuals engage more strongly in the task and feel good about it [20].

Based on the regulatory fit literature, we pose the following hypotheses for individuals with a high promotion focus (see Figure 2 for a summary of our hypotheses):

- **H4**: A gain-framed (vs. loss-framed) message for individuals with a high (vs. low) promotion focus leads to having a) the privacy calculus in favor of using the privacy-protection technology, b) a higher trust for the privacy-protection technology, and c) higher adoption of the privacy-protection technology.

Likewise, based on the regulatory fit literature, we hypothesize that the loss-framed message results in more positive attitudes and behaviors than the gain-framed messages for those with a high prevention focus:

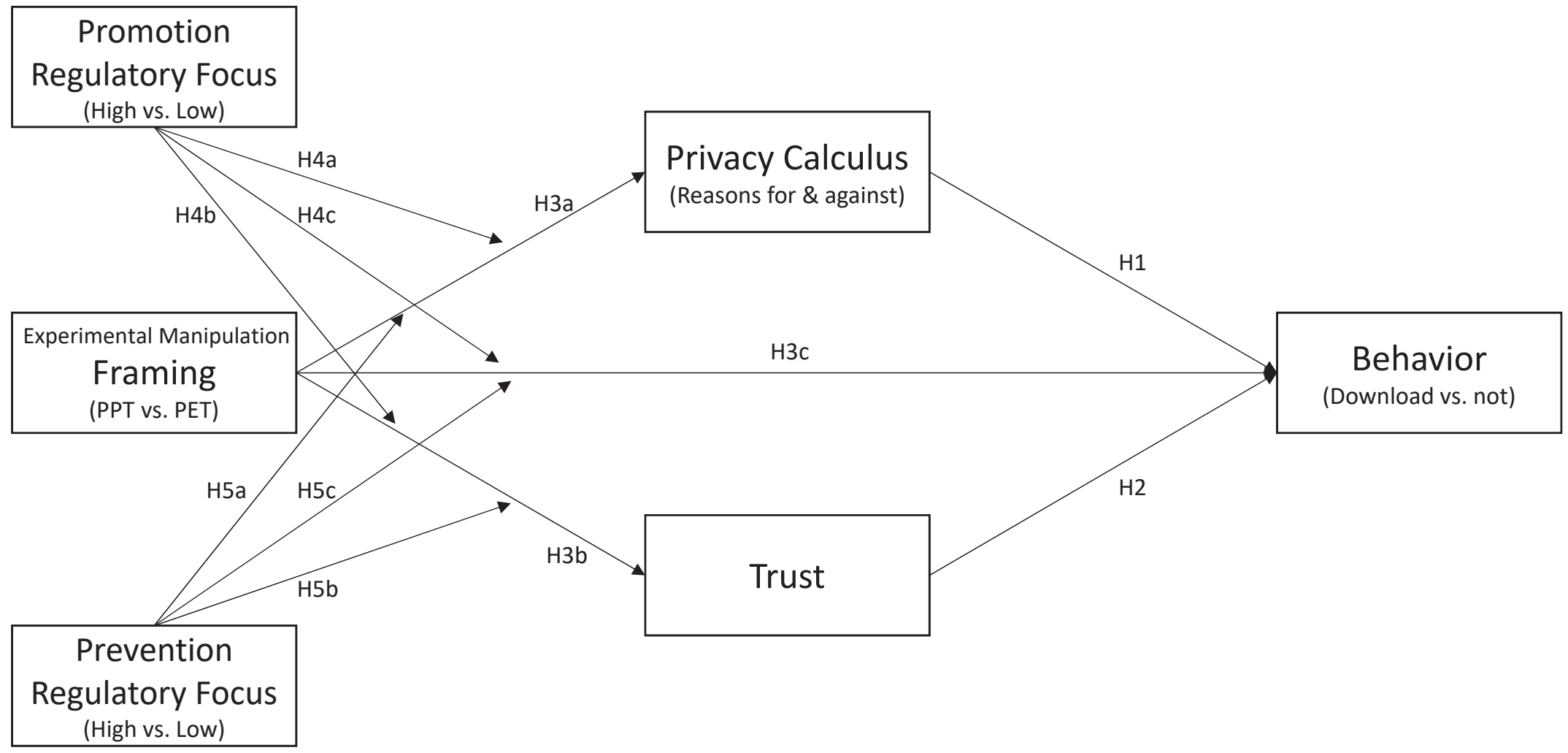

**Figure 2: The hypothesized model.**

- **H5**: A loss-framed (vs. gain-framed) message for individuals with a high (vs. low) prevention focus leads to having a) the privacy calculus in favor of using the privacy-protection technology, b) a higher trust for the privacy-protection technology, and c) higher adoption of the privacy-protection technology.

# 3 METHODS

## 3.1 Study Overview

We designed a between-subject experiment in which we present a privacy-protection technology—the IoT Inspector—either in a gain frame of *Privacy Enhancing Technology (PET)* or in a loss frame of *Privacy Preserving Technology (PPT)*. This study was reviewed and approved by our Institutional Review Board (IRB), as well as the institution of the developer team behind IoT Inspector, since the download of that tool was the subject of our study. After giving consent and agreeing to participate in the study, participants answered a brief survey measuring their privacy concerns as a control variable. Then, they read a short piece of information about the IoT Inspector (the framing manipulation) and answered post-survey questions, including their perception of IoT Inspector.We used Qualtrics to administer the survey. Finally, participants were given a chance to download and use the IoT Inspector. Whether they downloaded it or not is used as a binary indicator of their adoption of the IoT Inspector, the outcome variable in the hypothesized model. We validated downloads by giving those who proceeded to download the tool a unique ID to enter into the survey. Figure 3 shows the study overview. After finishing the study, participants received $5 as an incentive and were debriefed about the PET and PPT conditions and the purpose of the study.

## 3.2 Stimuli

IoT Inspector is an open-source software designed by researchers across several universities to help IoT device users monitor and understand the data-sharing practices of their home IoT devices [53]. It monitors the network traffic of IoT devices and allows users to track the frequency at which their smart devices send out data, the domains to which their data goes (e.g., Google.com), and the geographical location of these domains (e.g., USA). In order to design IoT Inspector's gain- and loss-framing introductory text, the authors had several meetings at which they discussed the text. The overall goal was to use the relevant gain (e.g., increase data security) or loss (e.g., decrease data breaches) terminology in each condition while keeping the text concise. One important criterion was that the framing manipulation should not have any semantic implications, such that the two PET and PPT versions should communicate the same information. This was especially important because if the two versions communicate different information, we would be unable to determine whether potential findings are due to using different gain vs. loss terminologies or due to the different information presented to the users. We present the full stimuli in the Appendices.

## 3.3 Measurement Instruments

We measured participants' baseline privacy concerns in the pre-survey as a control variable before presenting the framed text. Privacy concerns are the most studied variable in the privacy literature [16, 28], and are considered as an antecedent for adopting privacy and security technologies [10, 15]. We used Malhotra et al.'s Global Information Privacy Concerns [74]. This instrument includes five items (see Table 7). The responses were recorded on a 7-point Likert scale from "Strongly Disagree" to "Strongly Agree."

Then, we presented the framed text about the IoT Inspector, and measured participants' perceived trust in the tool, regulatory focus, cognitive aspect of decision-making through privacy calculus, technology use frequencies, and demographics. Perceived trust in the tool is a shortcut of privacy heuristic [66, 67]. McKnight's work suggests that trust has several dimensions [76]. We used the benevolence dimension, which is geared to measure the moral dimension of trust [111] and aligns better with the conceptual definition of trust that we discussed in Section 2.2.2. We tuned this construct to our context (e.g., "The IoT Inspector puts my interests first"; see Table 7 to view all statements). We used the Regulatory Focus Questionnaire (RFQ) developed by Higgins et al. [48] to measure participants' promotion and prevention regulatory foci (see Section 7). Following Higgins et al.'s [50] guidelines, we conducted a median split to identify individuals with high and low promotion and prevention regulatory focuses [1]. Similar to the pre-survey, we recorded responses on a 7-point Likert scale from "Strongly Disagree" to "Strongly Agree."

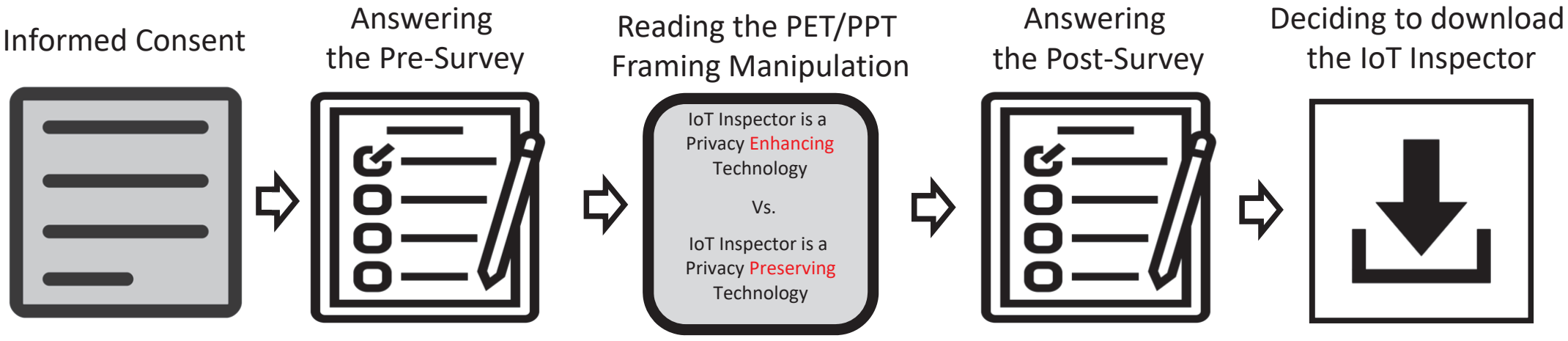

**Figure 3: An overview of our study.**

To capture the cognitively-mediated aspect of the decision to download the IoT Inspector (or not), we asked participants to list their reasons for or against using the tool. They had to type at least three and at most five reasons. Then, for each reason, they specified whether that reason was for or against using the tool. This method is a common means of process tracing in the psychology literature that helps scholars explore the cognitively-mediated aspect of decisions [59, 78]. It follows that the reasons that are listed first are often more important for people, and if people favor a choice, they tend to list more reasons *for* it rather than *against* it [59]. We considered these findings in coding the outcome of privacy calculus by summing the inverse signed ranks for these questions (please see below for the formula). Overall, a higher value means that participants have more positive reasons for using the IoT Inspector (i.e., the outcome of privacy calculus more greatly favors using the IoT Inspector).

$$\sum_{i=1}^{5} \frac{Q_i valence}{i} \tag{1}$$

In addition, participants answered questions about their ethnicity, age, and gender. We also measured the frequency of technology use with a question, "How frequently do you use technology (e.g., smartphone, internet)?" We recorded the responses on a 7-point Likert scale from "less often" to "almost constantly." Finally, participants were given a chance to download the tool. We clarified that the decision to download or not does not influence their incentives.

## 3.4 Participant Recruitment

Our power analysis showed that to identify a small effect (0.25) with a power of 0.95 and an $\alpha$ of 0.05, we need 210 participants. We recruited 238 US-based participants via Prolific, a crowd-sourcing platform (please see the Appendices to view the recruitment script). All participants agreed to participate in the study and were paid $5 through the Prolific platform after completing the study. We took several measures to ensure the quality of the data. First, we recruited participants with at least 90% successful approvals. These individuals are more likely to pay attention to the study. Furthermore, we included two attention-check questions in the survey to exclude those who may not read the survey carefully. In addition, to improve the ecological validity of our study, we used Prolific's built-in screening feature to recruit only those who use smart devices. This was important because the IoT Inspector would not be useful for those who do not use smart devices.

Of 238, two participants missed one or both attention check questions and were removed from the analysis. Thus, a total of 236 valid responses were collected. Participants were randomly assigned to the loss (N = 120) and gain (N = 116) conditions. We recruited participants with the goal of inclusivity, recruiting across a broad age range from 19 to 94 years (Mean = 42.86, SD = 19.03). One hundred twenty-five respondents identified as women, 104 identified as men, and seven as non-binary. One hundred eighty-one participants were White, 27 were Black or African American, eight were Asian, and 20 were multi-racial or people of varied ethnicities. Lastly, 21 participants had at least a master's degree, 90 had a four-year college degree, 89 had an associate's degree, 35 had a high school or an equivalent degree, and one had an educational level below high school.

## 3.5 Data Analysis

*3.5.1 Quantitative Analysis.* Although we borrowed measurement instruments from previously validated scales in the literature, we measured Cronbach's alpha to assess the reliability and internal consistency of measures in our context. The constructs we used in the survey showed a high internal consistency, with all of them having Cronbach's alpha values exceeding the acceptable thresholds of 0.7 [22, 82]. In addition, the measurement model showed good fit ($\chi^2(20)$ = 45.178, $p < 0.001$, RMSEA = 0.073, p = 0.086, CFI = 0.964, TLI = 0.949) [37, 52]. Consequently, we conducted a Structural Equation Model (SEM) to test our hypothesis. We used a robust maximum likelihood estimator, which is robust to non-normality [97, 113]. Besides the variables in the hypothesized model, we also included participants' reported privacy concerns in the SEM as a controlled variable. We conducted the SEM analyses in Mplus, and examined the explained variance (R-squared) in the outcome variable, download behavior. Additionally, we conducted difference testing using the Loglikelihoods [80] to study whether the model hypothesizing regulatory fit interactions is superior to the model without the two regulatory fit interaction effects. Finally, we analyzed a fully saturated model by including all possible two-way interaction effects to study the best possible model.

[1] While this is a common practice in regulatory focus literature and makes the results more comparable and easier to interpret, a median split may have statistical disadvantages. We analyzed the data after removing 30% of the sample around the medians and found that the effects of regulatory fit on trust and privacy calculus do not change. Therefore, we proceeded with the whole sample without removing data.

**Table 1: A breakdown of our participants' promotion and prevention regulatory focuses within each framing condition. The values in parentheses are the app downloads.**

| | Loss Framing (PPT) | | Gain Framing (PET) | |
|---|---|---|---|---|
| | High Prevention | Low Prevention | High Prevention | Low Prevention |
| High Promotion | 35 (15) | 26 (9) | 41 (12) | 27 (12) |
| Low Promotion | 27 (8) | 32 (4) | 18 (4) | 30 (6) |

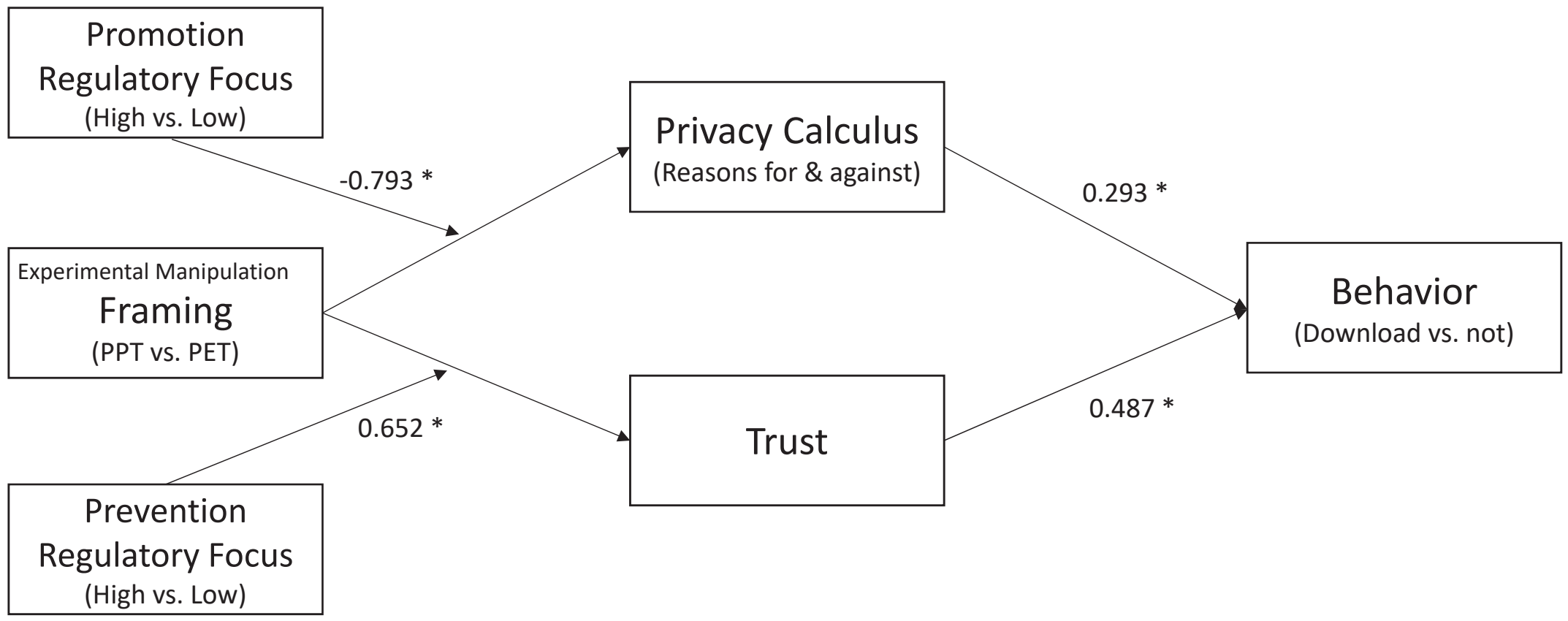

**Figure 4: We conducted an SEM with all the hypotheses as shown in Figure 2. To improve readability, we removed the non-significant findings from this figure. Table 2 reports all of the effects.**

*3.5.2 Qualitative Analysis.* At the end of the study, participants were asked to indicate their reasons for using or not using the IoT Inspector via an open-ended question. These responses were subjected to qualitative analysis using the six-stage thematic analysis approach [98]. Two researchers independently performed open coding using an inductive approach on de-identified, unlabeled data (i.e., participants' PET/PPT condition and regulatory foci were removed from the data) to mitigate potential biases. Both researchers then independently performed axial coding to develop initial categories. Next, both researchers reviewed the independently-developed categories and reached a consensus to define the final themes for the selective coding stage. Following selective coding, the frequencies of responses in each theme across the PET/PPT conditions and promotion and prevention regulatory focuses were reported.

## 4 RESULTS

In the following, we first report some descriptive statistics about participants' technology use and their attitudes toward the IoT Inspector. Then, we report the results of hypothesis testing, followed by our qualitative findings.

### 4.1 Descriptive Statistics

On average, participants reported using smart devices at least once a day, with 107 participants using them several times a day and 51 participants using them almost constantly. This shows that our participants are frequent technology users. In addition, they reported having an average of five smart devices. Therefore, the context of this study, smart home privacy, is relevant to these participants. Furthermore, participants had an average sum score of 22.800 for privacy concerns (min = 5, max = 35, SD = 5.982) and an average score of 15.444 for trust (min = 7, max = 21, SD = 2.257). In addition, participants listed a minimum of three and a maximum of five reasons *for* or *against* using the tool. On average, they entered 2.525 (SD = 1.497) reasons for and 1.182 (SD = 1.382) reasons against using the IoT Inspector. Therefore, they were more geared to list positive reasons than negative reasons. Across all participants, we collected 596 reasons for and 279 reasons against using the IoT Inspector. Ultimately, 70 participants downloaded the IoT Inspector. Therefore, the adoption rate was at about 30%. In addition, there were 121 participants with a high prevention focus and 129 participants with a high promotion focus. Table 1 shows a breakdown of individuals' regulatory focuses and the framing of the message they viewed.

### 4.2 Hypothesis testing

We tested the hypothesized model (see Figure 2) using a Structural Equation Modeling (SEM) framework. Overall, this model accounted for 23.2% of the variance in download behavior. Table 2 reports all of the SEM results. **Confirming H1**, we found a significant positive association between the privacy calculus and download behavior. When participants have more positive reasons for using the IoT Inspector, they are more likely to download the tool ($OR = 1.375, p = 0.026$). Furthermore, **we found support for H2**; by one standard deviation increase in trust, participants are 62.7% more likely to download the IoT Inspector ($p < 0.001$). However, we did not find a direct effect of framing on the download behavior ($p = 0.834$), privacy calculus ($p = 0.558$), or trust ($p = 0.231$, **H3a-c rejected**).

**Table 2: Results of the full SEM model. We used bold text to show the significant effects. Since the download behavior is a binary variable, we include both beta coefficients and odds ratios.**

| Variables | b (OR) | SE | p-value |
|---|---|---|---|
| *DV: Download Behavior* | R-squared = 22.6% | | |
| **Privacy Concerns** | **0.541 (1.718)** | **0.188** | **0.004** |
| **H1: ISR** | **0.319 (1.375)** | **0.128** | **0.013** |
| **H2: Triust** | **0.472 (1.603)** | **0.201** | **0.019** |
| H3c: Loss Framing (vs. Gain) | 0.069 (1.072) | 0.332 | 0.834 |
| High Promotion Focus (vs. Low) | 0.575 (1.777) | 0.333 | 0.084 |
| H4c: Loss Framing X High Promotion | 0.054 (1.056) | 0.654 | 0.934 |
| High Prevention Focus (vs. Low) | -0.103 (0.902) | 0.316 | 0.743 |
| H5c: Loss Framing X High Prevention | 0.664 (1.943) | 0.634 | 0.295 |
| *DV: Privacy Calculus* | R-squared = 5.3% | | |
| Privacy Concerns | -0.001 | 0.113 | 0.993 |
| H3a: Loss Framing (vs. Gain) | 0.113 | 0.194 | 0.558 |
| High Promotion Focus (vs. Low) | 0.226 | 0.188 | 0.229 |
| **H4a: Loss Framing X High Promotion** | **-0.958** | **0.378** | **0.011** |
| High Prevention Focus (vs. Low) | 0.351 | 0.184 | 0.057 |
| H5a: Loss Framing X High Prevention | 0.438 | 0.369 | 0.235 |
| *DV: Trust* | R-squared = 10.0% | | |
| Privacy Concerns | -0.028 | 0.088 | 0.753 |
| H3b: Loss Framing (vs. Gain) | -0.179 | 0.150 | 0.231 |
| **High Promotion Focus (vs. Low)** | **0.463** | **0.154** | **0.003** |
| H4b: Loss Framing X High Promotion | -0.293 | 0.295 | 0.321 |
| High Prevention Focus (vs. Low) | 0.266 | 0.151 | 0.077 |
| **H5b: Loss Framing X High Prevention** | **0.652** | **0.287** | **0.023** |

To explore the regulatory fit hypothesis, we study the interaction effect between individuals' regulatory foci and the framing conditions. A regulatory fit for participants with a high promotion was found to significantly change their privacy calculus, such that if they see the gain-framed message, they are more likely to have positive reasons for using the IoT Inspector ($b = 0.958, p = 0.011$, **H4a supported**[2]). However, regulatory fit for participants with a high promotion does not significantly improve trust ($p = 0.321$, **H4b rejected**), nor does it directly increase the likelihood of the download behavior ($p = 0.934$, **H4c rejected**).

A regulatory fit for participants with a high prevention focus involves the loss-framed message. When individuals with a high prevention focus see the privacy-preserving framing, they do not show a significantly different privacy calculus ($p = 0.235$, **H5a rejected**), but their trust perceptions are significantly higher by 0.625 standard deviations ($p = 0.023$, **H5b supported**). Furthermore, while a regulatory fit for those with a high prevention regulatory focus increases the likelihood of the download behavior by 94.3%, this effect is not significant ($p = 0.295$—**H5c rejected**). Lastly, to study whether the significant regulatory fit interaction effects improve the model, we compared the model with these interactions against the model without these interaction effects. The results show that adding such interaction effects significantly improves the model fit ($\chi^2(2) = 10.225$, $p = 0.006$). Table 3 summarizes the results of hypothesis testing.

[2]In Table 2 this effect has a negative sign (-0.958) as it shows the misfit situation of having Privacy Preserving framing for individuals who have high promotion regulatory focuses.

**Table 3: A summary of hypothesis testing results.**

| Hypotheses | Support |
|---|---|
| H1: Privacy Calculus | Supported |
| H2: Trust | Supported |
| H3: Framing | Not Supported |
| H4: Regulatory fit (for promotion) | Partially Supported |
| H5: Regulatory fit (for prevention) | Partially Supported |

## 4.3 Fully Saturated Model

In order to explore other possible effects, we studied a fully saturated model by adding all possible two-way interaction effects to the hypothesized model. We iteratively trimmed the newly added non-significant interaction terms and found only one additional significant interaction effect between privacy calculus and prevention regulatory focus, predicting download behavior ($OR = 1.833, p = .038$).

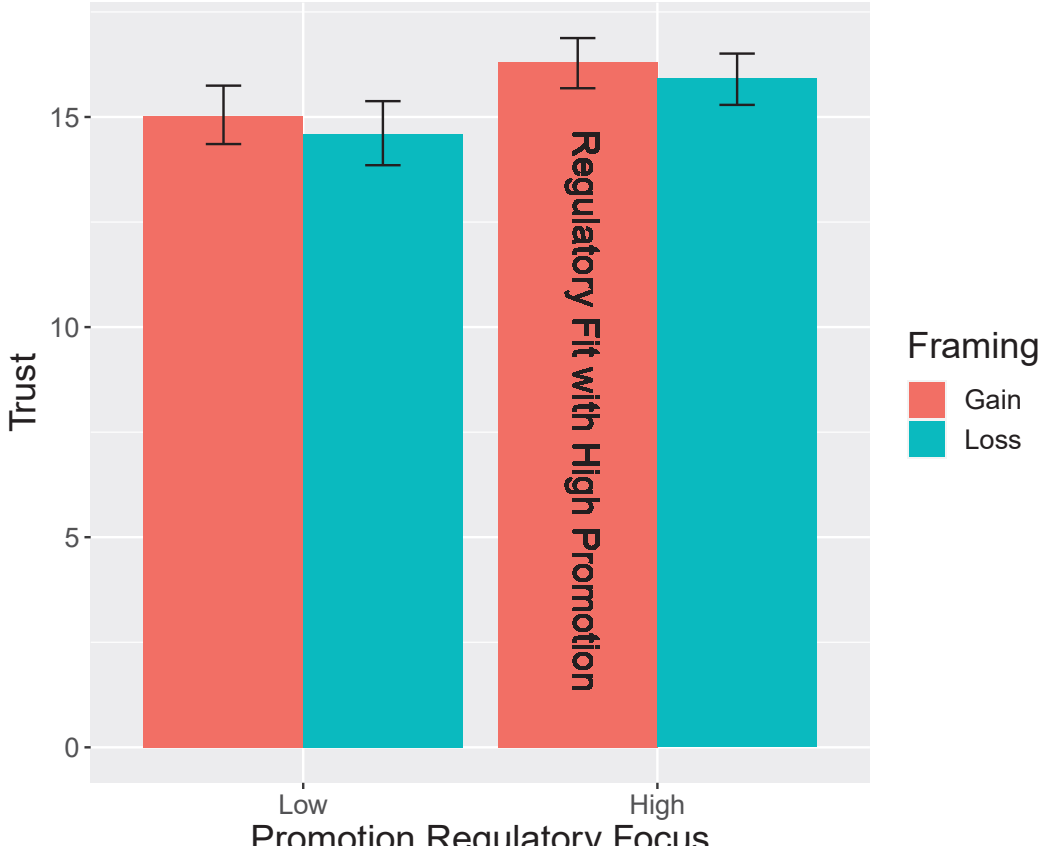

**(a) When participants with a high promotion focus experience regulatory fit (i.e., see the gain-framed message) they report the highest trust. However, this effect is not significant.**

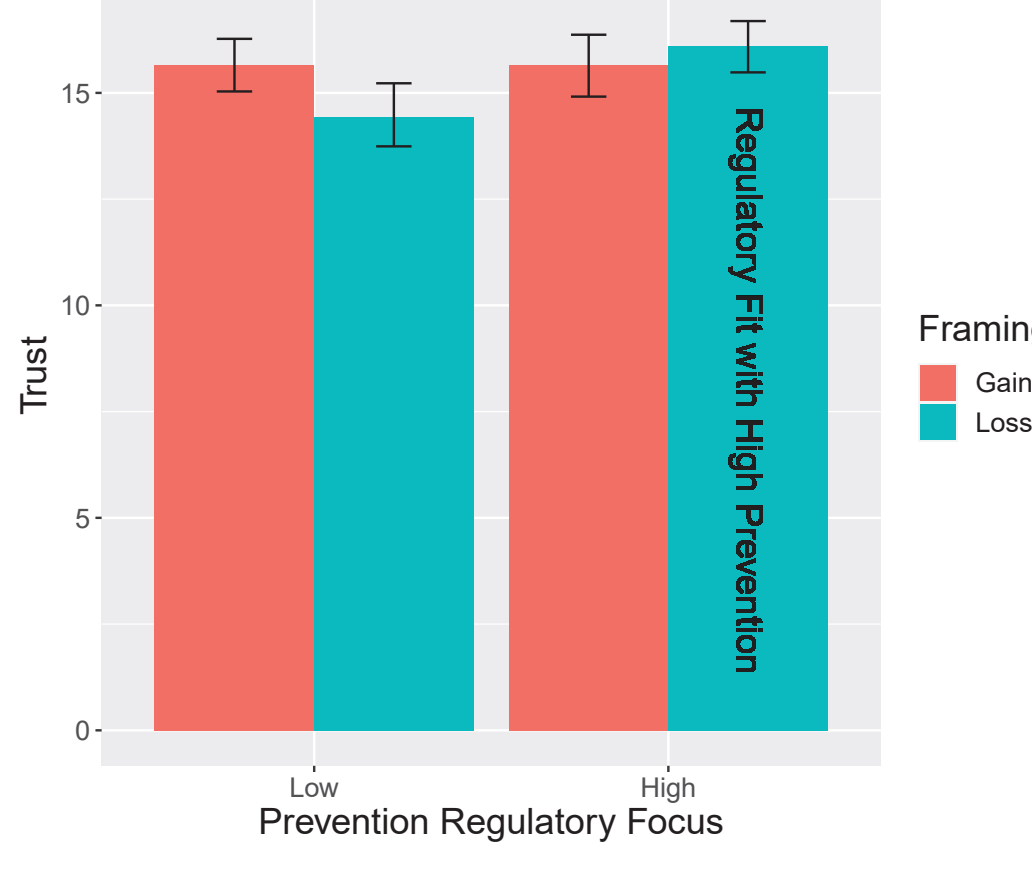

**(b) When participants with a high prevention focus experience regulatory fit (i.e., see the loss-framed message) they report the highest trust.**

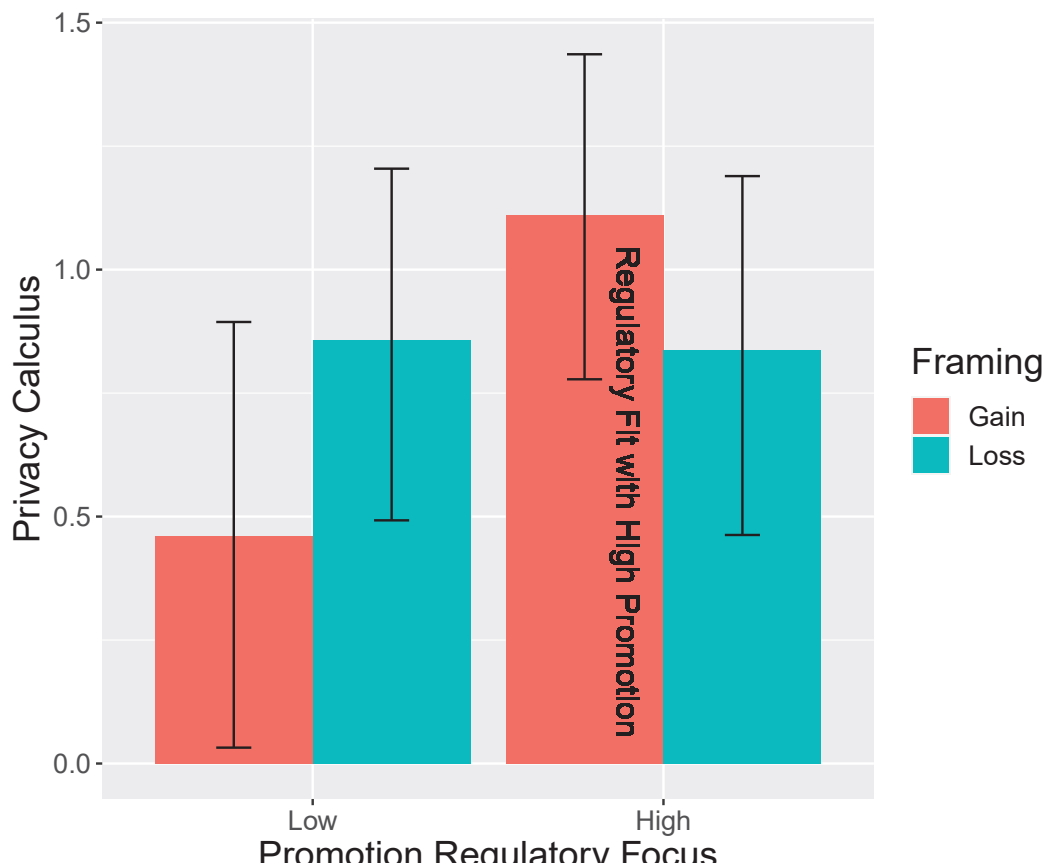

**(c) When participants with a high promotion focus experience regulatory fit (i.e., see the gain-framed message) they report the highest privacy calculus.**

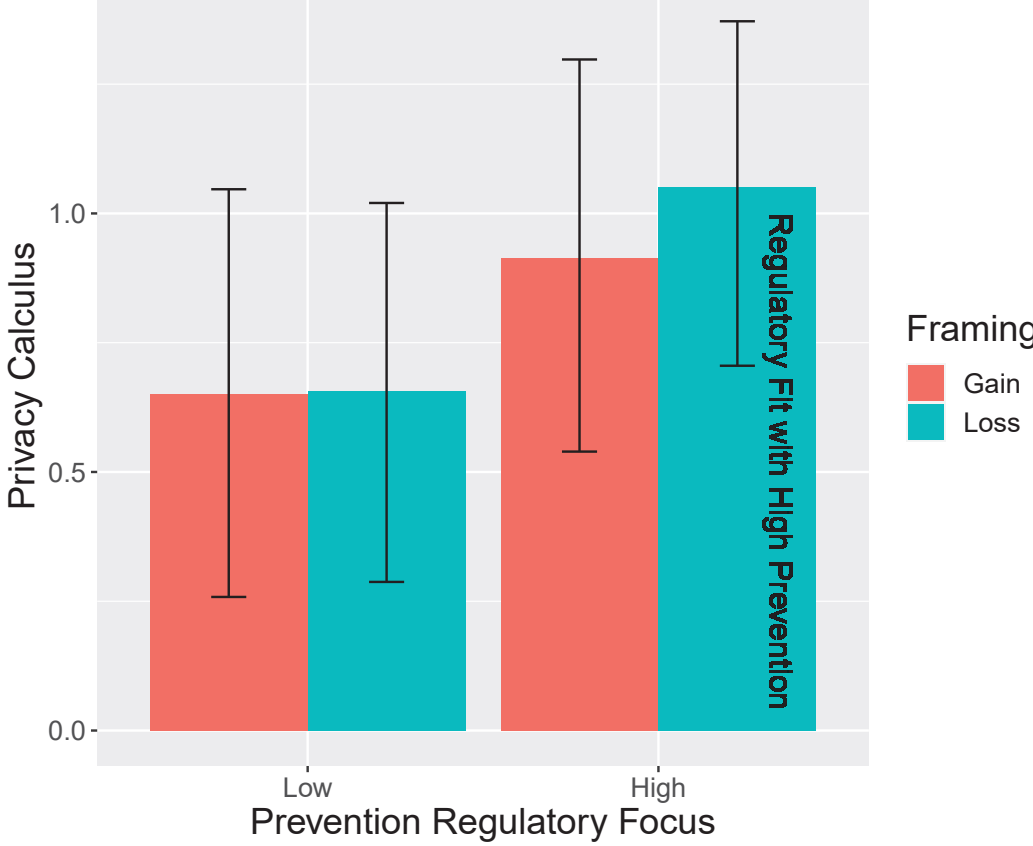

**(d) When participants with a high prevention focus experience regulatory fit (i.e., see the loss-framed message) they report the highest privacy calculus. However, this effect is not significant.**

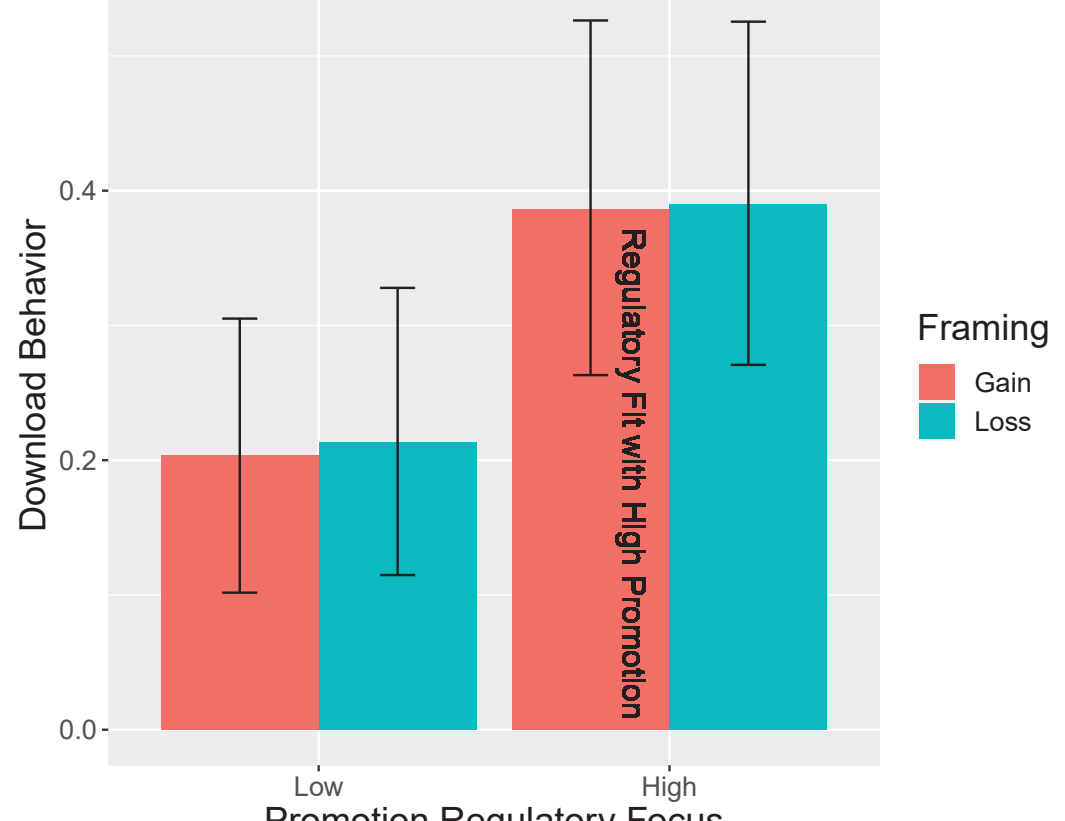

**(e) The direct effect of promotion regulatory fit is not significant.**

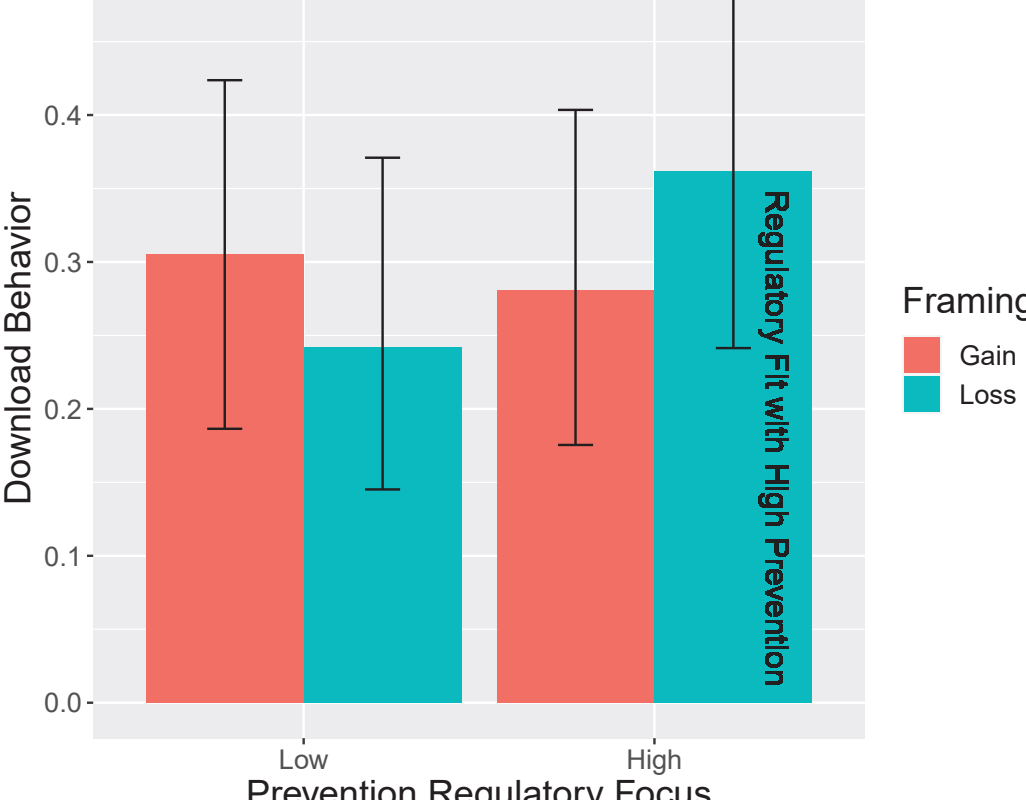

**(f) The direct effect of prevention regulatory fit is not significant.**

**Figure 5: While we only reached statistically significant effects across some of these graphs, regulatory fit conditions consistently lead to higher trust and privacy calculus.**

**Table 4: Direct and indirect effects of regulatory fit on dependent variables. "na" shows the paths that do not exist.**

| Regulatory fit | Trust | Privacy Calculus | Download Behavior |
|---|---|---|---|
| Promotion fit | Direct: 0.169, p = 0.320<br>Indirect: na<br>Total: 0.169, p = 0.320 | Direct: -0.163, p = 0.011<br>Indirect: na<br>Total: -0.163, p = 0.011 | Direct: -0.011, p = 0.878<br>Indirect: -0.060, p = 0.052<br>Total: - 0.072, p = 373 |
| Prevention fit | Direct: 0.155, p = 0.021<br>Indirect: na<br>Total: 0.155, p = 0.021 | Direct: 0.075, p = 0.234<br>Indirect:na<br>Total: 0.075, p = 0.234 | Direct: 0.058, p = 0.065<br>Indirect: 0.102, p = 0.179<br>Total: 0.160, p = 0.045 |

Chi-square difference testing suggests that adding this interaction term significantly improves the hypothesized model ($\chi^2(1) = 5.034$, $p = 0.024$). Finally, we report the direct and indirect effects of regulatory fit on dependent variables in Table 4.

## 4.4 Qualitative Insights

Thematic analysis of the reasons participants provided in favor or against using the IoT Inspector provides rich qualitative insights. Below, we present the major themes resulting from this analysis (see Table 5 for a summary).

*4.4.1 Themes Supporting Use of the IoT Inspector.* The most frequent theme for using the IoT Inspector (N = 423) was monitoring and protection. The second most prevalent theme in support of the IoT Inspector was that it could curtail users' efforts in managing their privacy (N = 53). Several statements showed an appreciation of how IoT Inspector can address users' worries and provide ease of mind (N = 43). Last, we observed several reasons for participant's appreciation of IoT Inspector as a tool that can improve transparency by helping users know what data is being collected (N = 23).

*4.4.2 Themes Opposing Use of the IoT Inspector.* Thematic analysis revealed that low trust was the most prevalent reason against using the IoT Inspector (N = 136). Fifty-seven reasons were listed against using the tool as it may entail spending extra time setting up or potentially require paid subscriptions. Several reasons suggested that some individuals were worried about the tool having compatibility issues with their existing tools (N = 35). Another theme that emerged as a reason for not using the IoT Inspector was the need for more information about the tool (N = 28). Nineteen reasons suggested a lack of concerns, as individuals were not disclosing important information online (see Table 5 for examples); 14 reasons argued that nothing could really protect one's privacy, and nine reasons focused on the need to hear other users' inputs, experiences, and reviews.

*4.4.3 Reflecting on Regulatory Fit Findings in Qualitative Themes.* In this section, we synthesize the qualitative findings with the quantitative results centered on regulatory fit. We highlight only the themes with a more substantial difference (e.g., not only by two or three frequencies). First, we explore patterns observed amongst individuals in the gain-framed condition who had a high promotion focus (regulatory fit with high promotion). Data monitoring and protection was the major listed reason in favor of using the IoT Inspector overall and was cited most (at 56.22%) among such individuals. In addition, we observed another noteworthy difference within the worry-mitigation theme, suggesting that these individuals consider IoT Inspector a means of mitigating their privacy worries (65.22%). However, these individuals do not appear to trust the tool differently than others. These results are in line with our quantitative findings, suggesting that the privacy calculus is more pronounced for regulatory fit with high promotion.

Prevention regulatory fit applies to individuals who read the loss-framed message and have a high prevention regulatory focus. While themes such as worry mitigation and reduced effort were not substantially different for people with high and low prevention regulatory foci in the PPT condition, low trust was the least cited reason for not using the IoT Inspector among the group with the high prevention focus (41.79%). This is in line with our quantitative findings, as it suggests that in the case of a prevention regulatory fit, individuals have a higher trust in the IoT Inspector.

# 5 DISCUSSION

Previous research explored the efficacy of persuasive messages based on regulatory fit in various domains (e.g., health [86], marketing [11]), but not in privacy. To the best of our knowledge, our work is the first to explore how regulatory fit can affect privacy-protection behaviors. Our findings shed light on the mechanisms through which regulatory fit influences persuasion in adopting a real-world privacy-protection technology.

We studied if and how privacy calculus and trust mediate the effects of regulatory fit on privacy decisions. We found that individuals with a high promotion focus who saw the gain-framed (PPT) message found the IoT Inspector more beneficial (i.e., their privacy calculus was more positive-leaning). High promotion focus is associated with approach orientation and commission bias (i.e., preference for action rather than inaction). With a gain-framed message, individuals with high promotion are more likely to be "motivated" to think about reasons "for" the action. However, since the total effect of promotion fit on the download behavior was not significant (see Table 4), we cannot conclude that a promotion fit actually led to a behavioral outcome. On the other hand, the total effect of prevention fit on the download behavior was significant. We found that individuals with a high prevention focus who saw the loss-framed (PET) message reported a higher level of trust in the IoT Inspector. Extant research has studied humans' "loss aversion," showing that losses trigger a heuristic mechanism in which individuals perceive losses as more significant than gains of similar size [58, 105]. It follows that loss-averse individuals tend to take more risks to avoid unfavorable outcomes [105]. Hence, a loss-framed message is more alarming to prevention-focused individuals, who inherently want to avoid losses. This leads them to heighten their trust in the IoT Inspector, through which they aim to avoid unfavorable outcomes.

**Table 5: The results of our qualitative analysis with themes and examples of each theme. The numbers represent the frequency of each theme in the reasons participants listed.**

| Themes | Examples | Frequency |
|---|---|---|
| *Reasons for using IoT Inspector* | | |
| Data monitoring and protection | "protect yourself from a data breach"<br>"It will monitor through info making sure nothing wrong is being shared around the web" | 423 |
| Reduce efforts | "It could save me time on monitoring my privacy"<br>"Helps make things easier" | 57 |
| Worry mitigation | "It makes you not worry much I feel it will look out for me"<br>"It brings me ease of mind in regards to smart devices" | 43 |
| Improved transparency | "IoT Inspector helps you monitor and understand how your devices interact with data..."<br>"Understanding what data is collected & why" | 23 |
| *Reasons against using IoT Inspector* | | |
| Low trust | "It feels like swapping one evil for another. IoT is going to tell me what other apps/sites are doing while doing it itself."<br>"Do not trust the IoT inspector" | 136 |
| Need information | "I need to better understand how IoT Inspector works"<br>"Do not have enough information about it" | 28 |
| Unconcerned (e.g., due to non-disclosure) | "I don't disclose barely any of my personal info "<br>"I'm not that worried about it because I haven't shared data that would hurt me" | 19 |
| Extra time, effort, or cost | "its just another thing I have to set up"<br>"If I have to pay for it, I really can't afford it" | 57 |
| Relinquish privacy | "There is nothing to entirely protect personal privacy data" | 14 |
| Need to see review | "Need to see reviews from real people" | 9 |
| Compatibility | "have to make sure my stuff runs fine through it" | 35 |

These results demonstrate that the mechanisms through which the persuading message influences users' adoption of privacy protection tools are complex, and the mediation analysis helped us scrutinize the adoption decision and gain deeper insights. This approach can inform research in other areas of persuasion. Researchers have examined the effectiveness of various persuading strategies, such as explaining the purpose of data collecting [62], informing users on the number of apps accessing their information [6], and indicating the level of security for different configuration options [116]. However, the efficacy of such persuading strategies is mixed, such that while some studies found the desired effects on persuasion [6, 19, 116], others did not find persuading messages as effective [8, 30, 54]. Studying the underlying mechanisms behind these effects and accounting for users' regulatory orientations can contribute to our understanding of the circumstances under which persuasive messages may or may not work (e.g., if the persuasive message does not align with an individual's regulatory focus, it may not work).

Our qualitative findings highlight the key reasons that potential users may consider when deciding to adopt a new technology, with implications for product design and marketing. While the vast majority of our participants highlighted the major application of the IoT Inspector (i.e., data monitoring and protection), many appreciated how the tool could save them time or mitigate their worries about their smart devices. Therefore, it is important for product designers to think not only about the immediate application of their product (e.g., data monitoring and protection) but also highlight and design for other relevant areas through which the product can benefit users (e.g., mitigating worries). Moreover, our results unveil several barriers to adopting IoT Inspector. We found trust to be the most profound barrier. While earning users' trust may take time, there may be some means to form an initial trust (e.g., through honest communication of the product's drawbacks [60]). Furthermore, while extensive information about a product may overwhelm some users [18], our results show that some users need more information before making the adoption decision. Therefore, it is important to make this information accessible to such users. However, addressing trust and providing information does not necessarily lead to adoption, as some users specified that they were unconcerned and simply did not need the tool.

In addition, our findings have important implications for policymakers who seek to promote responsible informed consent in policy design [92, 102]. Policymakers can leverage regulatory fit to promote privacy-oriented informed consent. For instance, a loss-framed message such as "To protect your privacy, please read the policy statement" may be more effective in engaging people with a high prevention focus than a gain-framed message of "To enhance your privacy, please read the policy statement," and the reverse may be true for people with a high protection focus. Further research should examine the efficacy of gain vs. loss framing in developing policy and regulation and determine if they are successful in motivating individuals with different regulatory foci.

**Table 6: Representation of each theme across experimental conditions and regulatory focuses.**

| Themes | High/Low Promotion Focuses | | | | High/Low Prevention Focuses | | | |
|---|---|---|---|---|---|---|---|---|
| | PET Framing | | PPT Framing | | PET Framing | | PPT Framing | |
| | High | Low | High | Low | High | Low | High | Low |
| *Reasons in favor of using IoT Inspector* | | | | | | | | |
| Data monitoring and protection | 122 (56.22%) | 95 (43.78%) | 104 (50.49%) | 102 (49.51%) | 111 (51.15%) | 106 (48.85%) | 106 (51.46%) | 100 (48.54%) |
| Reduce efforts | 15 (45.45%) | 18 (54.55%) | 7 (29.17%) | 17 (70.83%) | 18 (54.55%) | 15 (45.45%) | 13 (54.17%) | 11 (45.83%) |
| Worry mitigation | 15 (65.22%) | 8 (34.78%) | 10 (50%) | 10 (50%) | 15 (65.22%) | 8 (34.78%) | 11 (55%) | 9 (45%) |
| Improved transparency | 6 (60%) | 4 (40%) | 7 (53.85%) | 6 (46.15%) | 4 (40%) | 6 (60%) | 8 (61.54%) | 5 (38.46%) |
| *Reasons against using IoT Inspector* | | | | | | | | |
| Low trust | 35 (50.72%) | 34 (49.28%) | 34 (50.75%) | 33 (49.25%) | 32 (46.38%) | 37 (53.62%) | 28 (41.79%) | 39 (58.21%) |
| Need information | 12 (60%) | 8 (40%) | 5 (62.5%) | 3 (37.5%) | 8 (40%) | 12 (60%) | 4 (50%) | 4 (50%) |
| Unconcerned (e.g., due to non-disclosure) | 4 (25%) | 12 (75%) | 1 (33.33%) | 2 (66.67%) | 5 (31.25%) | 11 (68.75%) | 2 (66.67%) | 1 (33.33%) |
| Extra time, effort, or cost | 19 (50%) | 19 (50%) | 6 (31.58%) | 13 (68.42%) | 20 (52.63%) | 18 (47.37%) | 9 (47.37%) | 10 (52.63%) |
| Given up privacy | 1 (25%) | 3 (75%) | 3 (30%) | 7 (70%) | 2 (50%) | 2 (50%) | 5 (50%) | 5 (50%) |
| Need to see review | 4 (57.14%) | 3 (42.86%) | 2 (100%) | 0 (0%) | 4 (57.14%) | 3 (42.86%) | 0 (0%) | 2 (100%) |
| Compatibility | 4 (28.57%) | 10 (71.43%) | 9 (42.86%) | 12 (57.14%) | 6 (42.86%) | 8 (57.14%) | 7 (33.33%) | 14 (66.67%) |

However, policymakers, companies, and designers should also be cautious about the potential misuse of personalized persuasion. For example, dark pattern designers who attempt to maximize users' data disclosure [9, 14, 38] might use regulatory fit to amplify their data-collection efforts. As such, a dark pattern designer may frame a cookie consent request as a gain (e.g., "Allow cookies to make your shopping experience more convenient") for users with a promotion focus and as a loss (e.g., "Allow cookies so that we don't lose track of the items in your shopping cart") for users with a prevention focus. To do that, they need to learn their target audience's regulatory orientation. Overall, personalization may not be feasible without tracking any user data. While the 11-item RFQ is the main method currently used for inferring regulatory focus, there may be other means to discern regulatory orientation. For instance, IP addresses can unveil users' geographic location [104], and people living in countries with highly individualistic cultures are more loss-averse than collectivistic cultures [107, 110] and may be more likely to be promotion-focused. If such relationships are validated, dark pattern designers can infer individuals' regulatory orientations (e.g., based on geographic locations). Additionally, studies suggest that we can have assumptions about a user's regulatory focus based on demographics such as age (e.g., younger adults being more promotion-focused [70], and gender (e.g., women being more prevention-focused than men [42]). Similarly, prior research [61, 114] has shown that people's social media behavior can be used to predict their personality traits. Using similar methods, users' regulatory focus could potentially be elicited through their social media behavior and digital traces and make them susceptible to dark pattern interventions.

Finally, there is a need for more research and debate within the CHI community to explore the ethical boundaries around persuasion. While there is consensus on some applications of personalized persuasion being unethical [14, 84, 95], in certain other cases, there can be a fine line between persuasive design and manipulative design [93]. The HCI community, as one of the major user advocates, should establish a framework that sets ethical guidelines for using persuasive mechanisms in various contexts and decide whether or not, and to what extent, users' data can be used for personalization.

## 6 LIMITATIONS AND FUTURE WORK

We showed the efficacy of regulatory fit only in a limited context of adopting a privacy-protection technology. Future research can explore this effect in various privacy scenarios to study the generalizability of the findings (e.g., whether these personalized messages can motivate individuals to explore a new privacy feature in an existing app, read policy documents, or choose stronger passwords). In addition, we used a convenience sampling methodology by recruiting only US-based participants from Prolific. Prolific explicitly informs participants that they are recruited for participation in research [85] and requires participants to be paid a minimum hourly wage. Although Prolific provided high data quality in terms of attention, comprehension, and honesty [88], and prior studies found that Prolific participants were less dishonest and from a more diverse demographic than MTurkers [87], we face some limitations in terms of our recruiting approach. For example, participants in Prolific may not be interested in downloading an application as their main task is participating in the surveys. We notified participants that their download decision does not influence their participation reward, and about 30% of the participants downloaded the app. It is possible that this ratio would be different among non-prolific users. In addition, due to different cultural backgrounds or levels of digital literacy, it is possible to find different effects among non-prolific users. Overall, our results are not generalizable to the broader population, and our study is subject to other limitations, such as social desirability bias. Furthermore, we studied only participants' download behaviors and did not explore their actual usage of the IoT Inspector, nor did we follow up with participants to deeply understand their motives. Future research can longitudinally explore the regulatory fit and study if regulatory fit can have longitudinal effects (e.g., influence the duration and frequency of user interactions).

## 7 CONCLUSION

This study explored the effects of message personalization on adopting a privacy and security measure. More specifically, we communicated a privacy-protection technology using either a gain-framed message (Privacy Enhancing Technology) or a loss-framed message (Privacy Preserving Technology) to people with promotion and prevention regulatory foci. Our results suggest that individuals react to the same message differently based on their regulatory focus. Our study showed that a regulatory fit (i.e., tailoring the persuasive message to one's regulatory focus) can increase their trust and influence the outcome of their privacy calculus.

## ACKNOWLEDGMENTS

We would like to express our sincere gratitude to Professor Burcu Bulgurcu for her insightful and valuable feedback during the initial phase of this project.

## 8 APPENDICES

Below, we present the manipulation. There are several '/' symbols in the text. Participants in the PPT condition read the text before the '/', and participants in the PET condition read the text after the '/'.

> ***Preserve/Enhance*** *Your Privacy*
>
> *It is essential to* ***preserve/enhance*** *your privacy.* ***Privacy-Preserving Technologies (PPT) /Privacy-Enhancing Technologies (PET)*** *can help you* ***defend yourself/increase your protection*** *in the online world.* ***PPTs /PETs*** *are technologies that embody fundamental* ***privacy-preserving /privacy-enhancing*** *principles by* ***minimizing the disclosure/ maximizing the confidentiality*** *of your personal data and* ***decreasing data breaches/increasing data security***. ***PPTs /PETs*** *allow you to* ***protect/increase*** *the privacy of your personally identifiable information (PII) provided to and handled by services or applications.*
>
> *The smart devices in your home may potentially gather data without your knowledge, sometimes with malevolent intent. For instance, certain apps on your phone could potentially expose your data to harmful third parties. In response to this challenge, researchers at New York University have developed a tool called IoT Inspector.*

*The IoT Inspector is a privacy **preserving/enhancing** software designed to monitor the types of data being transmitted from your devices, such as audio, video, or text, and identify to which domains this information is being sent. By understanding the nature of the data each domain collects and the purpose of that specific domain, IoT Inspector can assess whether the collected data aligns with the domain's stated purpose. This allows IoT Inspector to flag any suspicious activity and **preserve/enhance** your privacy by **rejecting/accepting** network communications that are **unsafe/safe**. This way, IoT Inspector can **limit/increase** your SmartHome's **vulnerability/security**. IoT Inspector helps you **not lose/gain** control over your smart devices. The overall goal of IoT Inspector is for you to use your Smart Devices, and, at the same time, **avoid anxiety/gain peace of mind**.*

Recruitment Script:

*If you are a user of smart-home devices, please consider participating in our study. The study will take up to 15 minutes. You will receive a $5 compensation for participating in this study.*

**Table 7: In the regulatory focus questionnaire, items 1 (reversed), 3, 7, 9 (reversed), 10, and 11 (reversed) measure promotion regulatory focus and items 2 (reversed), 4 (reversed), 5, 6 (reversed), and 8 (reversed) measure prevention regulatory focus.**

| Privacy Concerns |
| --- |
| 1- All things considered, Smart Devices cause serious privacy problems. |
| 2- Compared to others, I am more sensitive about the way Smart Devices handle my personal information |
| 3- To me, it is the most important thing to keep my privacy intact from Smart Devices. |
| 4- I believe other people are too concerned with Smart Devices' privacy issues. |
| 5- I am concerned about threats to my personal privacy today. |
| **Trust** |
| 1- This IoT Inspector puts my interests first. |
| 2- This IoT Inspector keeps my interests in its mind. |
| 3- This IoT Inspector wants to understand my needs and preferences. |
| **Regulatory Focus Questionnaire** |
| 1- Compared to most people, I am typically unable to get what you want out of life. |
| 2- Growing up, I "crossed the line" by doing things that my parents would not tolerate. |
| 3- I accomplished things that got me "psyched" to work even harder. |
| 4- I often got on my parents' nerves when I was growing up. |
| 5- I often obeyed the rules and regulations that were established by my parents. |
| 6- Growing up, I acted in ways that my parents thought were objectionable. |
| 7- I often do well at different things that I try. |
| 8- Not being careful enough has gotten me into trouble at times. |
| 9- When it comes to achieving things that are important to me, I find that I don't perform as well as I ideally would like to do. |
| 10- I feel like I have made progress toward being successful in my life. |
| 11- I have found very few hobbies or activities in my life that capture my interest or motivate me to put effort into them. |